\documentclass{amsart}
\usepackage{amssymb,stmaryrd}
\usepackage{amsfonts}
\usepackage{amstext}
\usepackage{algorithmic}
\usepackage{algorithm}
\usepackage{graphicx}
\usepackage{epstopdf}
\usepackage[all]{xy}
\usepackage{MnSymbol}

\numberwithin{equation}{section}

\theoremstyle{definition}

\theoremstyle{remark}

\newcommand{\R}{{\mathbb{R}}}

\newcommand{\C}{{\mathbb{C}}}
\newcommand{\Z}{{\mathbb{Z}}}

\newcommand{\<}{{\langle}}

\renewcommand{\>}{{\rangle}}

\newcommand{\wedgeq}{{\wedge\kern-5pt\cdot}}

\newcommand{\tens}{\otimes}

\newcommand{\id}{{\rm id}}

\newcommand{\extd}{{\rm d}}
\newcommand{\del}{{\partial}}

\begin{document}

\title{Quantum gravity on a square graph}


\author{Shahn Majid}
\address{Queen Mary, University of London\\
School of Mathematics, Mile End Rd, London E1 4NS, UK}

\email{s.majid@qmul.ac.uk}


\begin{abstract} We consider functional-integral quantisation of the moduli of all  quantum metrics defined as square-lengths $a$ on the edges of a Lorentzian square graph. We determine correlation functions and find a fixed relative uncertainty $\Delta a/\<a\>= 1/\sqrt{8}$ for the edge square-lengths relative to their expected value $\<a\>$. The expected value of the geometry is a rectangle where parallel edges have the same square-length. We compare with the simpler theory of a quantum scalar field on such a rectangular background. We also look at quantum metric fluctuations relative to a rectangular background, a theory which is finite and interpolates between the scalar theory and the fully fluctuating theory. 
\end{abstract}
\maketitle 

\section{Introduction}

The quantum spacetime hypothesis -- the idea that the coordinates of spacetime are better modelled as noncommutative operators as an expression of quantum gravity effects was proposed in \cite{Ma:pla} on the basis of position-momentum reciprocity. The possibility itself was speculated on since the early days of quantum theory, while the specific argument in \cite{Ma:pla} was that the quantum phase space of some part of quantum gravity that contains position and momentum is obviously noncommutative, but its division into position and momentum is arbitrary and in particular should be interchangeable. Since there is generically gravity with curvature on spacetime then there should also generically be curvature in momentum space. But under quantum-group Fourier transform in the simplest cases, this is equivalent to noncommutative position space. Thus the search for quantum groups  at the time that were both noncommutative and `curved' (noncocommutative) became a toy model of the search for quantum gravity. The resulting `bicrossproduct quantum groups' later resurfaced as quantum Poincar\'e groups for action models such  $[x_i,t]=\lambda x_i$, the Majid-Ruegg model \cite{MaRue} notable for its testable predictions via quantum Fourier transform\cite{AmeMa}.  Other early models were the Dopplicher-Fredenhagen-Roberts one \cite{DFR} adapting work of Snyder\cite{Sny} and  the $q$-Minkowski one, see \cite{Ma:book}.  Quantum spacetimes and position-momentum duality are also visible in 3D quantum gravity\cite{MaSch} e.g. as a curved $S^3$ momentum space with the angular momentum algebra $U(su_2)$ as quantum spacetime\cite{BatMa}, as first proposed by t' Hooft\cite{Hoo} for other reasons. There were also several models of quantum field theory on flat quantum spacetimes, such as \cite{Gro,FL}, and indications from loop quantum gravity\cite{Ash}. 

Some 25 years later, however, we have a much better idea what spacetime (and quantum phase space) that is both noncommutative and curved actually means mathematically. Specifically, we will use a constructive {\em quantum Riemannian geometry} formalism \cite{BegMa1,BegMa2,BegMa3,BegMa4,DV1,DV2,Mou,Mad, Ma:gra,MaPac,MaTao}  coming in part out of experience with the geometry of quantum groups  but not limited to them. This starts with a differential calculus and quantum metric, in contrast to the more well-known approach  to noncommutative geometry of Connes starting with a `Dirac operator' (spectral triple)\cite{Con}, though not necessarily incompatible\cite{BegMa3}. Recent lectures notes for the constructive formalism are in \cite{Ma:ltcc} and a brief outline is in  Section~\ref{secoutline}. 
Our own motivation for this effort was that quantum Riemannian geometry should then be a much better starting point from on which to build quantum gravity as it already includes quantum gravity corrections.  It is therefore a fair question as to whether, now, one can actually build models of quantum gravity using this more general conception of Riemannian geometry.

In this paper we give an example of such a model, albeit a baby one with only four points. We will not need the full machinery of quantum Riemannian geometry (but it is important that it exists so that what we do is not too ad-hoc). In fact our algebra $A$ will be commutative, namely functions on a discrete set $X$ (the four points in our example), but differential forms will necessarily not commute with functions, so we still need quantum Riemannian geometry. A limited version of the formalism for this case is in the preliminary Section~\ref{secX}.  The key point is that quantum differential structures in this context are given by directed graphs with $X$ as the vertex set, and a metric is given by nonzero real numbers attached to the directed edges. It is tempting to think of these `edge weights' as lengths but in fact a better interpretation (e.g. from thinking about the graph Laplacian) is as the square of metric lengths. Note also that there are potentially two such square-lengths for every edge but we restrict to the `edge-symmetric' case where the two weights are required to be the same. The quantum Riemannian geometry of a quadrilateral or `square graph' in this context was recently solved in \cite{Ma:eme} and is recalled in Section~\ref{secquad} along with a small  extension (Case~2) from \cite{BegMa:book}. We adapt these directly for a `Lorentzian square graph' where horizontal edge weights will be taken negative. We will generally  refer to the square-length of an edge as the magnitude of the number associated to an edge, and the geometric  timelike or spacelike length is the square root of this. 

Section~\ref{secscalar} contains the first new results, including a warm-up quantisation of a massive scalar free field on a Lorentzian rectangular background (where parallel edges have the same length  and non-parallel edges are orthogonal) in Section~\ref{secscalarconstant}. We also cover the scalar theory on a set of just two points one edge in Section~\ref{scalarZ2} and  some results for a general curved non-rectangular background in Section~\ref{scalarcurved}.  Next, we come in Section~\ref{secqg}  to our model of quantum gravity, covering both a full quantisation of the edge square-lengths (which washes out much of the structure) and the intermediate quantisation of edge-length fluctuations relative to a rectangular background. In the former, the correlation functions turn out to be real and to imply a uniform relative uncertainty of $1/\sqrt{8}$ for all edge square-lengths as well as that the expectation values for parallel edges are the same. The intermediate theory gives a similar picture in the limit where the background edge-length goes to zero. It also asymptotes for large background edge lengths to a product of two massless scalar theories, each on 2 points (namely an edge and its parallel edge regarded now as vertices for the scalar theory). The paper concludes with some directions for this model needed further to develop the interpretation. Finite quantum gravity has been considered in the past in Connes spectral triple approach, notably \cite{Hale}, but without directly comparable results due to the different starting point. Our approach also has a very different character and methodology from current lattice quantum gravity. 

 For the quantisation, we focus on the physical case of $\imath$ in the action, albeit the Euclideanised case could also be of interest.  We work in units where  $\hbar=c=1$ and have one coupling constant $\beta$ in the action for the scalar case and another $G$ for the metric case.   In fact, standard dimension-counting does not apply in the model which is in some sense 0-dimensional (4 points) and in another sense 2-dimensional (there is a top form of degree 2 and a 2-dimensional cotangent bundle).
 
\section{Preliminaries: formalism of quantum Riemannian geometry}

Here we recall briefly some elements the constructive `bottom up' approach to quantum Riemannian geometry particularly as used in \cite{BegMa1,BegMa2,BegMa3,BegMa4,Ma:gra,MaTao}  that we will use.  An introduction to the formalism is in  \cite{Ma:ltcc} and  for discrete sets in \cite{Ma:gra}, with the square mainly from \cite{Ma:eme}. 

\subsection{Bimodule approach} \label{secoutline} We will not need to full generality of the theory and give only the bare bones for orientation. It is important, however, that it exists so that our discrete geometry will be part of a functorial construction and not ad-hoc to the extent possible. We give only the bare bones for orientation purposes. 

Thus, one can replace a classical space equivalently by a suitable algebra of functions on the space and the idea now to allow this to be any algebra $A$ with identity (so no actual space need exist). We replace the notion of differential structure on a space but specifying a bimodule $\Omega^1$ of differential forms over $A$. A bimodule means we can multiply a `1-form' $\omega\in\Omega^1$ by `functions' $a,b\in A$ either from the left or the right and the two should associate according to $(a\omega)b=a(\omega b)$. We also need $\extd:A\to \Omega^1$ as `exterior derivative' obeying reasonable axioms the most important of which is the Leibniz rule. We require $\Omega^1$ to then extend to forms of higher degree giving a graded algebra  $\Omega$ with $\extd$ obeying a graded-Leibniz rule with respect to the graded product $\wedge$ and $\extd^2=0$. 

Next, on an algebra with differential we define a metric as an element $g\in \Omega^1\tens_A\Omega^1$ which is invertible in the sense of a map $(\ ,\ ):\Omega^1\tens_A\Omega^1\to A$ which commutes with the product by $A$ from the left or right and inverts $g$. We also usually require quantum symmetry in the form $\wedge(g)=0$. 

Finally, we need we need a connecton. A left connection on $\Omega^1$ is a map $\nabla :\Omega^1\to \Omega^1\tens_A\Omega^1$ obeying a left-Leibniz rule $\nabla(a\omega)=\extd a\tens \omega+ a\nabla \omega$ and this is a {\em bimodule connection}\cite{DV1,DV2,Mou,Mad,BegMa1} if there also exists a bimodule map $\sigma$ such that
\[ \sigma:\Omega^1\tens_A\Omega^1\to \Omega^1\tens_A\Omega^1,\quad \nabla(\omega a)=(\nabla\omega)a+\sigma(\omega\tens\extd a).\]
The map $\sigma$ if it exists is unique, so this is not additional data but a property that some connections have. A connection might seem mysterious but if we think of a map $X:\Omega^1\to A$ that commutes with the right action by $A$ as a `vector field' then we can evaluate $\nabla$ as a covariant derivative $\nabla_X=(X\tens\id)\nabla:\Omega^1\to \Omega^1$ which classically is then a usual covariant derivative. Bimodule connections extend automatically to tensor products as $\nabla(\omega\tens\eta)=\nabla\omega\tens\eta+\sigma(\omega\tens\id\tens\id)\nabla\eta$ so that metric compatibility makes sense as $\nabla g=0$. A connection is called  QLC or `quantum Levi-Civita' if it is  metric compatible and the torsion also vanishes, which in our language amounts to $\wedge\nabla=\extd$ as equality of maps $\Omega^1\to \Omega^2$. We also have a Riemannian curvature $R_\nabla:\Omega^1\to \Omega^2\tens\Omega^1$. Ricci requires more data and the current state of the art (but probably not the only way) is to introduce a lifting map $i:\Omega^2\to\Omega^1\tens\Omega^1$. Applying this to the left output of $R_\nabla$ we are then free to `contract' by using the metric and inverse metric to define ${\rm Ricci}\in \Omega^1\tens_A\Omega^1$ \cite{BegMa2}. The Ricci scalar is then $S=(\ ,\ ){\rm Ricci}\in A$.  More canonically, we have a geometric quantum Laplacian $\Delta=(\ ,\ )\nabla\extd: A\to A$ defined again along lines that generalise the classical concept to any algebra with differential structure, metric and connection.  

Finally, and critical for physics, are unitarity or `reality' properties. We work over $\C$ but assume that $A$ is a $*$-algebra (real functions, classically, would  be the self-adjoint elements). We require this to extend to $\Omega$ as a graded-anti-involution (so reversing order and extra signs according to the degree of the differential forms involved) and to commute with $\extd$. `Reality' of the metric comes down to $g^\dagger=g$ and of the connection to $\nabla\circ *= \sigma\circ\dagger\circ \nabla$, where $(\omega \tens_A\eta)^\dagger=\eta^*\tens_A \omega^*$ is the $*$-operation on $\Omega^1\tens_A\Omega^1$ \cite{BegMa1,BegMa2}. These reality conditions in a self-adjoint basis (if one exists) and in the classical case would ensure that the metric and connection coefficients  are real.

\subsection{Quantum Riemannian geometry of a single edge} \label{secX}

We will be interested in the case of  $X$ a discrete set and $A=\C(X)$ the usual commutative algebra of complex functions on it as our `spacetime algebra'. It is an old result that all possible 1-choices of $(\Omega^1,\extd)$ are in 1-1 correspondence with directed graphs with $X$ as the set of vertices. Here $\Omega^1$ has basis $\{\omega_{x\to y}\}$ over $\C$ labelled by the arrows of the graph and differential $\extd f=\sum_{x\to y}(f(y)-f(x))\omega_{x\to y}$. In this context a quantum metric\cite{Ma:gra}
\[ g=\sum_{x\to y}g_{x\to y}\omega_{x\to y}\tens\omega_{y\to x}\in \Omega^1\tens_{C(X)}\Omega^1\]
requires weights $g_{x\to y}\in \R\setminus\{0\}$ for every edge and for every edge to be bi-directed (so there is an arrow in both directions).  The calculus over $\C$ is compatible with complex conjugation on functions $f^*(x)=\overline{f(x)}$ and $\omega_{x\to y}^*=-\omega_{y\to x}$. 

Finding a QLC for a metric depends on how $\Omega^2$ is defined and one case where there is a canonical choice of this is $X$ a group and the graph a Cayley graph generated by right translation by a set of generators. Here the edges are of the form $x\to xi$ where $i$ is from the generating set and the product is the group product. In this case there is a natural basis of left-invariant 1-forms $e_i=\sum_{x\to xi} \omega_{x\to xi}$. These obey the simple rules 
\[ e_i f= R_i(f)e_i,\quad \extd f=\sum_i (\del_i f)e_i,\quad \del_i=R_i-\id,\quad R_i(f)(x)=f(xi)\]
defined by the right translation operators $R_i$ as stated. Moreover, in this case  $\Omega$ is generated by the $e_i$ with certain `braided-anticommutation relations' cf. \cite{Wor}. In the case of an Abelian group this is just the usual  Grassmann algebra on the $e_i$ (they anticommute). 

For $X=\Z_2$ and its unique generator the Cayley graph is $0 \leftrightarrow 1$ with just one edge with two arrows. There is one invariant form $e=\omega_{0\to 1}+\omega_{1\to 0}$ with $e f= \tilde f e$, $\extd f=(\del f)e$ where $\tilde f$ swaps the values at $0,1$ and $\del f=(\tilde f-f)$. We have $e^2=0$ and $e^*=-e$ for the $*$-exterior algebra and metric
\[ g=a e=a(0)\omega_{0\to 1}\tens\omega_{1\to 0}+a(1)\omega_{1\to 0}\tens\omega_{0\to 1}\]
with 2 non-vanishing real parameters $a(0)=g_{0\to 1}$ and $a(1)=g_{1\to 0}$ as the two arrow weights. The {\em edge-symmetric} case of a single weight associated to either direction is $a(1)=a(0)$ or $\del a=0$. The inverse metric is $(e,e)=1/\tilde a$. A short calculation shows that there exists a QLC if and only if $\rho:=\tilde a/a = \pm 1$, i.e. $a(1)=\pm a(0)$. This has the form 
\[ \nabla e=b e\tens e,\quad b(0)=1-q, \quad b(1)=1-q^{-1}\rho,\quad |q|=1\]
with one circle parameter $q$ (the restriction here is for a $*$-preserving connection). The connection necessarily has zero curvature and has geometric Laplacian 
\begin{equation}\label{LapZ2} \Delta f=(\ ,\ )\nabla\extd f=(\ ,\ )\nabla(\del f e)=(\ ,\ )(\del^2 f +(\del f)b)(e\tens e)=-(\del f)\left ({q\over \bar a}+{q^{-1}\over a}\right)\end{equation}
which makes sense for any $a$ but as mentioned only comes from $\nabla$ a QLC if $a(1)=\pm a(0)$. The natural choice is the + case so that the metric is edge-symmetric (a single real number associated to the edge) and $\Delta f=(\del f)(q+q^{-1})/a$ then has real as opposed to imaginary eigenvalues. (In the other case, we would only have $q=\pm 1$ with $\Delta f=0$ for real coefficients.) We proceed in the edge-symmetric case. Then  the scalar action for a free massive field in 1 time and 0 space dimensions is,
\begin{equation}\label{actionZ2} S_f=\sum_{\Z_2} \mu f^*(\Delta+m^2)f=(q+q^{-1})|f(1)-f(0)|^2+ a m^2\left(|f(0)|^2+|f(1)|^2\right)\end{equation}
where we see that edge weight $a$ has the square of length dimension so that $am^2$ is dimensionless. We used $\mu=a>0$ as the constant  `measure' in the sum since the edge is viewed is the time-like direction at least when $q=1$ (but we are not limited to this.)

\subsection{Quantum Riemannian geometry of a quadrilateral}\label{secquad}

\begin{figure}\[ 
\includegraphics[scale=.65]{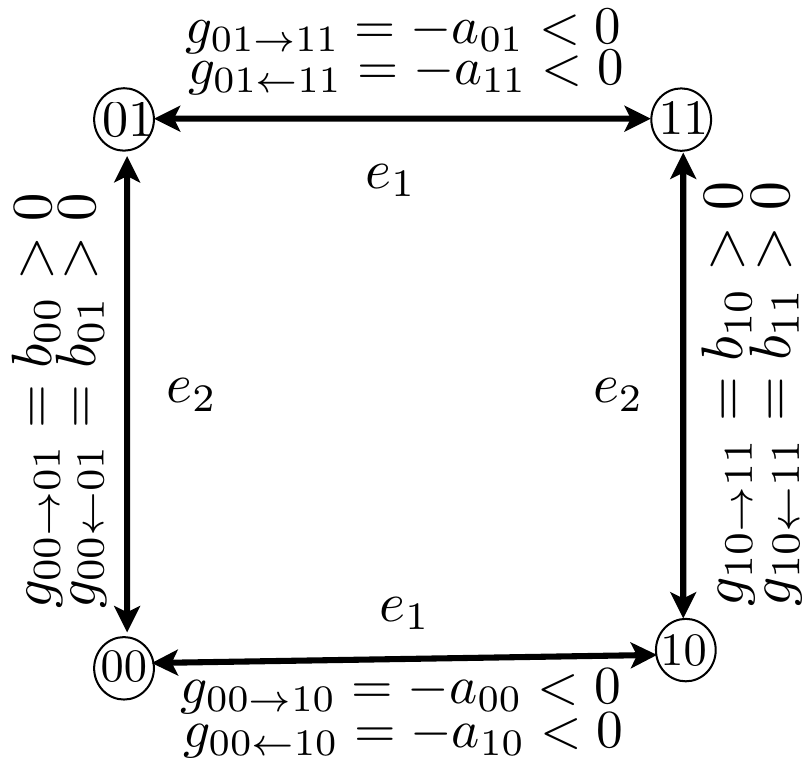}\]
\caption{Metric coefficients defined by functions $a,b>0$ are interpreted as arrow square-lengths on a Lorentzian square graph with $e_1$ spacelike. \label{fig1}}\end{figure}

We now consider $X=\Z_2\times\Z_2$ with its canonical 2D calculus given by a square graph with vertices $00,01,10,11$ in an abbreviated notation as shown in Figure~\ref{fig1}. This is  Cayley graph with generators $10,01$ and correspondingly two basic 1-forms 
\[ e_1=\omega_{00\to 10}+\omega_{01\to 11}+\omega_{10\to 00}+\omega_{11\to 01},\quad e_2=\omega_{00\to 01}+\omega_{10\to 11}+\omega_{01\to 00}+\omega_{11\to 10}.\] 
Now the relations $e_i f=(R_if)e_i$ and partials $\del_i=R_i-\id$ are defined by  $R_1$ that shifts by 1 mod 2 (i.e. takes the other point) in the first coordinate, similarly for $R_2$ in the second. The $*$-exterior algebra is the usual Grassmann algebra on the $e_i$ (they anticommute) with $e_i^*=-e_i$. The general form of a quantum  metric and its inverse are 
\[  g=-a e_1\tens e_1+b e_2\tens e_2,\quad (e_1,e_1)=-{1\over R_1 a},\quad (e_2,e_2)={1\over R_2 b},\quad (e_1,e_2)=(e_2,e_1)=0\]
where the coefficients are functions $a,b>0$ to reflect a Lorentzian signature with $e_1$ the spacelike direction. In terms of the graph, their 8 values are equivalent to the values of $g$ on the 8 arrows as shown in Figure~\ref{fig1}, where $a_{ij}=a(i,j)$ is a shorthand and similarly for $b_{ij}$. As for the $\Z_2$ case above, it is natural to focus on the edge-symmetric case where the edge weight assigned to an edge does not depend on the direction of the arrow. This means $\del^1 a=\del^2b=0$ and we assume this now for simplicity. 

\subsubsection*{Case 1: Generic metric QLCs}

For generic (non-constant) metrics the QLCs were recently found \cite{Ma:eme} and we adapt this with $a\to -a$ in present conventions. There is a 1-parameter family of QLCs 
\[ \nabla e_1=(1+Q^{-1})e_1\tens e_1+(1-\alpha)(e_1\tens e_2+e_2\tens e_1)+ {b\over a}(R_2\beta-1)e_2\tens e_2,\]
\[ \nabla e_2={a\over b}(R_1\alpha-1)e_1\tens e_1+(1-\beta)(e_1\tens e_2+e_2\tens e_1)+(1-Q)e_2\tens e_2,\]
where $Q,\alpha,\beta$ are functions on the group defined as
\begin{equation}\label{Q} Q=(q,q^{-1},q^{-1},q),\quad \alpha=({a_{01}\over a_{00}}, 1, 1, {a_{00}\over a_{01}}),\quad \beta=(1, {b_{10}\over b_{00}},  {b_{00}\over b_{10}},1)\end{equation}
when we list the values on the points in the above vertex order. Here $q$ is a free parameter and we need $|q|=1$ for a $*$-preserving connection.

The Riemann curvature has the general form $R_\nabla e_i=\rho_{ij}e_1\wedge e_2\tens e_j$ where \cite{Ma:eme}
\begin{align*} \rho_{11}=&Q^{-1}R_1\alpha-Q\alpha+(1-\alpha)(R_1\beta-1)+{R_2 a\over a}(R_2\beta-1)(R_2R_1\alpha-1)\\
\rho_{12}&=Q^{-1}(1-\alpha)+\alpha(R_2\alpha-1)-Q^{-1}{R_1b\over a}(\beta^{-1}-1))-{b\over a}(R_2\beta-1)R_2\beta\end{align*}
and similar formulae for $\rho_{2i}$. If we use the obvious antisymmetric lift  $i(e_1\wedge e_2)={1\over 2}(e_1\tens e_2-e_2\tens e_1)$  then 
\[ {\rm Ricci}=((\ ,\ )\tens\id)(\id\tens i\tens\id)(\id\tens R_\nabla)(g)={1\over 2}\begin{pmatrix}-R_2\rho_{21} & -R_2\rho_{22}\\ R_1\rho_{11} & R_1\rho_{12}\end{pmatrix}\]
as the matrix of coefficients on the left in our tensor product basis. Applying $(\ ,\ )$, the resulting Ricci scalar curvature is 
\[ S={1\over 2 }\left(   {R_2\rho_{21}\over a}+{R_1 \rho_{12}\over b}\right)={1\over 4 ab}\left((3+q-(1-q)\chi){\del_2 a\over\alpha}+(1-q^{-1}-(3+q^{-1})\chi) {\del_1 b\over\beta}\right)\]
where $\chi=(1,-1,-1,1)$. Next, we take  $\mu=ab>0$ in our conventions and
\begin{equation}\label{EH} S_g=\sum_{\Z_2\times\Z_2} \mu S=(a_{00}-a_{01})^2({1\over a_{00}}+{1\over a_{01}})-(b_{00}-b_{10})^2({1\over b_{00}}+{1\over b_{10}})\end{equation}
independently of $q$. If we had taken  the Euclidean signature as in \cite{Ma:eme} then both terms would enter with $+$ and the minimum would be zero, for the constant or rectangular case. 

Finally, the geometric Laplacian for the generic metric solutions comes out as\cite{Ma:eme}
\begin{eqnarray}\label{Lapgen} \Delta f=(\ ,\ )\nabla(\del_i f e_i)=-\left({Q^{-1}-R_2\beta\over a}\right)\del_1 f- \left({Q+R_1\alpha\over b}\right)\del_2 f\end{eqnarray}
again with a change of sign for $a$.

\subsubsection*{Case 2: Constant metric QLCs}

It is not central to the paper but we mention that in the `rectangular' case where  $a,b$ are constant, so $\alpha=\beta=1$, there is a much larger  4-parameter moduli of QLCs given in \cite{BegMa:book} by $P=(p_1,p_2^{-1},p_1^{-1},p_2)$ and $Q=(q_1,q_1^{-1},q_2^{-1},q_2)$ for nonzero constants $p_i,q_i$, where as before we list the values on the points in order $00,01,10,11$. The connections and curvature take the form\cite{BegMa:book}
\[ \nabla e_1=(1-P)e_1\tens e_1,\quad \nabla e_2=(1-Q)e_2\tens e_2,\]
\[ R_\nabla e_1=(\del_2 P)e_1\wedge e_2\tens e_1,\quad R_\nabla e_2=-(\del_1 Q)e_1\wedge e_2\tens e_2,\]
with the connection is $*$-preserving when  $|p_i|=|q_i|=1$. So the moduli space of QLCs here is  the 4-torus $T^4$. The Ricci tensor for the same antisymmetric lift as before now gives
\[ {\rm Ricci}={1\over 2}\left((\del_2 P^{-1})e_2\tens e_1-(\del_1 Q^{-1})e_1\tens e_2\right),\quad S=0.\]
The Laplacian for the 4-parameter QLCs for constant $a,b$ is 
\begin{eqnarray}\label{Lapcon} \Delta f=\left({P+1\over a}\right)\del_1 f- \left({Q+1\over b}\right)\del_2 f\end{eqnarray}
again with change of sign of $a$ compared to \cite{BegMa:book}.  The moduli of QLCs for generic metrics in Case 1 reduces when the metric is constant to the special case $P=-Q^{-1}$ with $q_1=q_2$. There may also in principle be intermediate QLCs when just one of $a$ or $b$ is non-constant.

\section{Quantisation of free scalar fields on two and four points}\label{secscalar}

We will adopt a `functional integral' approach where we parametrize the fields and integrate
an action over all fields. It is convenient to work in momentum space where our fields are Fourier transformed on the underlying finite group. 

\subsection{Scalar field on a single edge} \label{scalarZ2}

For simplicity, we take $f$ real-valued (the  complex case has the same form of free field action for the real and imaginary components separately). Furthermore, we expand
\[ f= f_0+ f_1\phi;\quad \phi(i)=(-1)^i\]
for Fourier coefficients $f_i$ on $\Z_2=\{0,1\}$. Then $f(0)=f_0+f_1$ and $f(1)=f_0-f_1$ and
\[ S_f=4(q+q^{-1}) f_1^2 + 2 a m^2 (f_0^2+f_1^2).\]
The path integral $Z=2\int\extd f_0\extd f_1e^{\imath S_f}$ has a Gaussian form which we compute as usual using
\[ Z_\alpha=\int_{-\infty}^\infty \extd k\, e^{\imath\alpha k^2}=\sqrt{\pi\over\alpha} e^{\imath\pi\over 4}\]
which implies
\begin{equation}\label{vevk2scalar} \< k^2\>:={\int_{-\infty}^\infty \extd k\, e^{\imath\alpha k^2} k^2\over \int_{-\infty}^\infty \extd k\, e^{\imath\alpha k^2}}={1\over \imath Z_\alpha}{\del\over\del\alpha}Z_\alpha={\imath \over 2\alpha}\end{equation}
and hence in our case the correlation functions
\[ \< f(0)f(1)\>=\<f(1)f(0)\>=\<f_0^2-f_1^2\>={\imath \over 4}\left( {1\over  a m^2}-{1\over   a m^2+2(q+q^{-1})}\right)\]
\[ \< f(0)f(0)\>= \<f(1)f(1)\>=\<f_0^2+f_1^2\>={\imath \over 4}\left( {1\over  a m^2}+{1\over   a m^2+2(q+q^{-1})}\right)\]
where $\<f_0f_1\>=0$ as each integrand is then odd. There is an infra-red divergence as expected as $m\to 0$ but in the massive case there are no divergences and hence no renormalisation needed until we consider interactions.

\subsection{Scalar fields on a Lorentzian rectangle}\label{secscalarconstant}

We start with the constant metric case so $a,b>0$ are constant  horizontal and vertical edge lengths (with the former negative in terms of edge weight),  and we work in `momentum space' with Fourier modes 
\[ 1,\quad \phi(i,j)=(-1)^i=(1,1,-1,-1),\quad \psi(i,j)=(-1)^j=(1,-1,1,-1),\quad \phi\psi=\chi\]
\[\del_1\phi=-2\phi,\quad \del_2\phi=0,\quad \del_1\psi=0,\quad \del_2\psi=-2\psi,\quad \del_1\chi=\del_2\chi=-2\chi. \]
Thus,  we let 
\[ f= f_0+ f_1\phi+ f_2 \psi + f_3 \chi\]
for the plane wave expansion of a general scalar field. As before, we focus on the real-valued case so the $f_i$ are real. We are mainly interested in this paper in generic metrics so we start with the specialisation to the rectangle of the  Laplacian (\ref{Lapgen}) for the generic QLCs for this, with their circle parameter $q$ and corresponding function $Q$ which we  expand as
\[ Q={1\over 2}\left(q+q^{-1}+(q-q^{-1})\chi\right),\quad Q^{-1}={1\over 2}\left(q+q^{-1}-(q-q^{-1})\chi\right).\]
Then
\[ \Delta f=2{Q^{-1}-1\over a}(f_1 \phi + f_3 \chi)+ 2{Q+1\over b} (f_2 \psi + f_3 \chi)\]
\[   \Delta1=0,\quad  \Delta \phi={q+q^{-1}-2\over a}\phi-{q-q^{-1}\over a}\psi,\quad  \Delta\psi={q+q^{-1}+2\over b}\psi+{q-q^{-1}\over b}\phi\]
\[\Delta\chi=(q-q^{-1})\left(-{1\over a}+{1\over b}\right)+\left({q+q^{-1}-2\over a}+{q+q^{-1}+2\over b}   \right)\chi\]
has one zero mode $1$, one mode built from $1,\chi$ with real eigenvalue read off from the bottom line and two more modes which are linear combinations of $\phi,\psi$ which one can show also have real eigenvalues for $a,b>0$. Next, we use the action
$S_f=\sum_{\Z_2\times\Z_2} \mu f(\Delta +m^2) f$  and we take $\mu=ab>0$. Then \begin{align*} S_f
&=4\big(a {(q+1)^2\over q}(f_2^2+f_3^2)+b {(q-1)^2\over q}(f_1^2+f_3^2)+(q-q^{-1})(a-b)(f_1f_2+f_0f_3)\\ &\quad + m^2 ab\left(f_0^2+f_1^2+f_2^2+f_3^2\right)\big).\end{align*}
The action again has real coefficients if and only if $q=\pm 1$ so that the $q-q^{-1}$ term vanishes. This is also the case where this class of QLCs with the rectangular metric has zero quantum Riemann curvature, $R_\nabla=0$. We now focus on this case where the interpretation is clearer. Then 
\[ S_f=\begin{cases}16 a (f_2^2+f_3^2) & {\rm if}\ q=1\\  - 16 b (f_1^2+f_3^2) & {\rm if}\ q=-1 \end{cases}+ 4 m^2 ab\left(f_0^2+f_1^2+f_2^2+f_3^2\right)\]
This is now in diagonal form so we can immediately write down the functional integral quantisation using
\[ Z=\int\extd f_0\extd f_1\extd f_2\extd f_3 e^{{\imath\over \beta} S_f}\]
and regarding the $a,b,m$ as parameters. We have inserted a coupling constant $\beta$ in the exponential of square length dimension for book-keeping. All four integrals have the same Gaussian form as the $\Z_2$ case, so from (\ref{vevk2scalar}) we can compute 2-point functions for $q=1$ that
\begin{align*} \<f_{00}f_{01}\>&=\<f_{10}f_{11}\>=\<f_0^2+f_1^2-f_2^2-f_3^2\>={\imath\beta\over 4}\left({1\over a b m^2}-{1\over ab m^2+4a}\right)\\
 \<f_{00}f_{10}\>&=\<f_{01}f_{11}\>=\<f_0^2-f_1^2+f_2^2-f_3^2\>=0 \\
 \< f_{00}f_{11}\>&=\<f_{01}f_{10}\>=\<f_0^2-f_1^2-f_2^2+f_3^2\>=0\\
\<f_{ij}^2\>=\<f_0^2&+f_3^2+f_1^2+f_2^2\>={\imath\beta\over 4}\left({1\over a b m^2}+{1\over ab m^2+4a}\right)\end{align*}
where we use the shorthand $f_{ij}=f(i,j)$ so that $f_{00}=f_0+f_1+f_2+f_3$, $f_{01}=f_0+f_1-f_2-f_3$ and $f_{00}f_{01}=(f_0+f_1)^2-(f_2+f_3)^2$, etc. As before, the cross terms do not contribute  due to the parity of the integrands. The result for $q=-1$ is similar,
\begin{align*} \<f_{00}f_{01}\>&=\<f_{10}f_{11}\>=\<f_0^2+f_1^2-f_2^2-f_3^2\>=0\\
\<f_{00}f_{10}\>&=\<f_{01}f_{11}\>=\<f_0^2-f_1^2+f_2^2-f_3^2\>={\imath\beta\over 4}\left({1\over a b m^2}-{1\over ab m^2-4b}\right)\\
 \< f_{00}f_{11}\>&=\<f_{01}f_{10}\>=\<f_0^2-f_1^2-f_2^2+f_3^2\>=0\\
\<f_{ij}^2\>=\<f_0^2&+f_3^2+f_1^2+f_2^2\>={\imath\beta\over 4}\left({1\over a b m^2}+{1\over ab m^2-4b}\right).\end{align*} 

Finally, because we are in the rectangular metric case, the quantum Riemannian geometry actually admits a larger moduli of QLCs with Laplacian (\ref{Lapcon}) where
\[ \Delta f=- 2{P+1\over a} (f_1 \phi + f_3 \chi)+ 2{Q+1\over b} (f_2 \phi + f_3 \chi).\]
The $P,Q$ depend on four modulus 1 parameters $p_i,q_i$ and a similar analysis to the above gives the action has real coefficients if and only if $p_i,q_i$ have values $\pm 1$ or $P,Q$ are chosen from $\pm1,\pm \chi$. For example, $P=Q=1$ has
\[ \Delta f={2\over a}\del_1 f- {2\over b}\del_2 f=-{4\over a}(f_1 \phi + f_3 \chi)+ {4\over b}(f_2 \phi + f_3 \chi).\]
Hence $\Delta1=0$,  $\Delta \phi=-{4\over a}\phi$,   $\Delta\psi={4\over b}\psi$ and $\Delta\chi=(-{4\over a}+{4\over b})\chi$ gives us the eigenmodes modes, with just one zero mode. We also have action
\[ S_f=\sum_{\Z_2\times\Z_2} \mu f(\Delta +m^2) f=-16 b(f_1^2+ f_3^2)+16 a(f_2^2+f_3^2) + 4 a b m^2(f_0^2+f_1^2+f_2^2+f_3^2)\]
for a massive free field with `measure' $\mu=ab$. This again has diagonal form which is a  composite of our previous $q=\pm1$ cases. Then we can immediately write down  the 2-point functions as 
\begin{align*} \<f_{00}f_{01}\>&=\<f_{10}f_{11}\>={\imath\beta\over 8}\left({1\over a b m^2}+{1\over ab m^2 - 4b}-{1\over ab m^2+4a}-{1\over ab m^2+4a-4b}\right)\\
 \<f_{00}f_{10}\>&=\<f_{01}f_{11}\>={\imath\beta\over 8}\left({1\over a b m^2}-{1\over ab m^2 - 4b}+{1\over ab m^2+4a}-{1\over ab m^2+4a-4b}\right) \\
 \< f_{00}f_{11}\>&=\<f_{01}f_{10}\>={\imath\beta\over 8}\left({1\over a b m^2}-{1\over ab m^2 - 4b}-{1\over ab m^2+4a}+{1\over ab m^2+4a-4b}\right)\\
\<f_{ij}^2\>&={\imath\beta\over 8}\left({1\over a b m^2}+{1\over ab m^2 - 4b}+{1\over ab m^2+4a}+{1\over ab m^2+4a-4b}\right)\end{align*}
where $f_{i,j}=f(i,j)$.  As before, the massless case of the above would have an infra-red divergence, here regularised by the mass parameter $m$. 

\subsection{Scalar field on a curved  non-rectangular background}\label{scalarcurved} 

Here we briefly consider the general case of a scalar field with a general (generically curved) non-rectangular edge-symmetric metric. In this case, it is convenient to also Fourier expand the metric in terms of four real momentum-space coefficients as
\begin{equation}\label{abfou} a=k_0+k_1\psi,\quad b=l_0+l_1\phi\end{equation}
\begin{equation}\label{fouab} a_{00}=k_0+k_1,\quad a_{01}=k_0-k_1,\quad b_{00}=l_0+l_1,\quad b_{10}=l_0-l_1.\end{equation}
So the preceding section was $k_1=l_1=0$ and $a=k_0, b=l_0$ while more generally $k_0,l_0>0$ are each the average of two parallel edge square-lengths (with the actual horizontal metric edge weights being negative) and $k_1,l_1$ are the amount of fluctuation. We restrict to $|k_1|<k_0$ and $|l_1|<l_0$ in order that our metric does not change signature. In either case it is useful to change variables from $k_1,l_1$ to the {\em relative fluctuations}  $k=k_1/k_0$ and $l=l_1/l_0$ both in the interval $(-1,1)$. As before, we keep the scalar field real valued for simplicity (the complex case is entirely similar). 

We need the 1-parameter QLCs for the general metric, with a modulus one parameter $q$ and Laplacian (\ref{Lapgen}), resulting in $S_f=\sum\mu f(\Delta +m^2)f$  given by
\begin{align*} S_f&=\sum_{\Z_2\times\Z_2}\mu f(\Delta+m^2)f=4\big(k_0 {(q+1)^2\over q}(f_2^2+f_3^2)+l_0 {(q-1)^2\over q}(f_1^2+f_3^2)\\
&+(q+q^{-1})\left(k k_0 (f_0f_2+f_1 f_3)+l l_0 (f_0 f_1+f_2f_3)\right)\\
&+(q-q^{-1})(k_0-l_0)(f_1f_2+f_0f_3)- 2 l l_0 \left(q -q^{-1}-2\right)f_1f_3+2 k k_0 \left(q -q^{-1}+2\right)f_2 f_3\\
&+m^2 k_0l_0\left(f_0^2+f_1^2+f_2^2+f_3^2+2l(f_0f_1+f_2f_3)+2k(f_0f_2+f_1f_3)+2k l(f_1f_2+f_0f_3)\right)\big)\end{align*}
As a check, this is invariant under the interchange 
\begin{equation} q\ \leftrightarrow\ -q^{-1};\quad k_0\ \leftrightarrow\ - l_0;\quad k\ \leftrightarrow\ l;\quad f_1\ \leftrightarrow\ f_2.\end{equation}
In the Euclidean square graph version (before we changed $a$ to $-a$) we have $k_0\leftrightarrow l_0$ and the symmetry reflects the ability to interchange the horizontal and vertical directions of the square, but note that we also have to change $q$. A similar symmetry was noted for the eigenvalues of $\Delta$ in \cite{Ma:eme} but the above is more  relevant since it accounts also for the `measure' $\mu$ in the action. 

On the other hand, we again need $q=\pm1$ for the action to have real coefficients (to kill the $q-q^{-1})$ term) and, without loss of generality, we focus on the case $q=1$; the other case is similar given the symmetry mentioned above. In this case
\begin{align*}S_f&=4\big(4 k_0   (f_2^2+f_3^2)+2 k k_0 (f_0f_2+f_1 f_3)+2 l l_0 (f_0 f_1+f_2f_3)+ 4 l l_0 f_1f_3+4 k k_0  f_2 f_3\\
&+m^2 k_0l_0\left(f_0^2+f_1^2+f_2^2+f_3^2+2l(f_0f_1+f_2f_3)+2k(f_0f_2+f_1f_3)+2k l(f_1f_2+f_0f_3)\right)\big).\end{align*}

Moreover, the action is quadratic in the $f_i$ so the functional integration is that of a Gaussian, with the result that the partition function for a free field can be treated in just the same way as the in the case of a rectangular background in Section~\ref{secscalarconstant}, after diagonalisation of the quadratic form underlying $S_f$. At issue for this are the eigenvalues of this quadratic form. Its trace is 
\[ 8(k_0 {(q+1)^2\over q}+l_0 {(q-1)^2\over q}+ 2 m^2 k_0 l_0)=16k_0(2+  m^2 l_0)\]
when $q=1$. Thus the sum of the eigenvalues (even for complex $q$) is real but the eigenvalues themselves for generic values are complex unless $q=\pm1$, when they are real. They are also generically but not necessarily nonzero (this is reasonable where there is curvature). For example, in the massless case with $q=1$ the determinant of the underlying quadratic form is 
\[ 256 (k^2 k_0^2 - 4 k_0 l l_0 - l^2 l_0^2) (k^2 k_0^2 + 4 k_0 l l_0 - 
   l^2 l_0^2)\]
so that there are four 3-surfaces in the four-dimensional metric moduli space where an eigenvalue  vanishes (e.g. giving $l$ in terms of $k,k_0,l_0$).  

In short, the two real choices $q=\pm1$ each behave similarly to the rectangular background case although the exact eigenvalues and hence the correlation functions depend in a complicated way on the background metric.

\section{Quantised metric on a quadrilateral}\label{secqg}

We now consider quantisation of the general edge-symmetric metric.   Again it is convenient to use the Fourier mode expansion as given at the start of Section~\ref{scalarcurved} where $k_0,l_0>0$ are the  average horizontal and vertical square-lengths respectively (the actual horizontal edge weights are negative) and $k=k_1/k_0,l=l_1/l_0$ are the relative fluctuations. Then the Einstein-Hilbert action (\ref{EH}) becomes
\begin{equation}\label{actionkl} S_g=\sum_{\Z_2\times\Z_2}\mu S=k_0\alpha(k)- l_0\alpha(l);\quad \alpha(k)=  {8 k^2\over 1-k^2}\end{equation}
in our Lorentzian signature case. This has square-length dimension needing a coupling constant, which we call $G$, of square-length dimension. 

\subsection{Full quantisation} We functionally integrate over all edge square-lengths with our given Lorentzian signature. Under our change of variables, the measure of integration becomes $\extd a_{00}\extd a_{01}\extd b_{00}\extd b_{10}=4\extd k_0\extd k_1\extd l_0\extd l_1=4  \extd k_0 \extd l_0 \extd k \extd l\, k_0 l_0$ and the partition function becomes 
\begin{align*} Z&= 2\int_{-1}^1\extd k\int_0^L \extd k_0 k_0  e^{{\imath \over G} k_0 \alpha(k)}= 4 G^2 \int_0^1\extd k{\extd\over\extd\alpha}|_{\alpha=\alpha(k)}{1-e^{ {\imath L\over G}\alpha}\over \alpha}=4G^2\int_0^\infty\extd \alpha{\extd k\over\extd\alpha}{\extd\over\extd\alpha}\left({1-e^{ {\imath L\over G}\alpha}\over \alpha}\right)\end{align*}
for the $k_0,k$ integration, times its complex conjugate  for the $l_0,l$ integration. Here  we regularised an infinity by limiting the $k_0$ integral to $0\le k_0\le L$ rather than allowing this to be unbounded. We also noted that $\alpha(k)$ is an even function and monotonic in the range $k\in [0,1)$, hence in this range we changed variable to regard $k= \sqrt{\alpha\over 8+\alpha}$ as a function of $\alpha\in [0,\infty)$. For fixed $L$ the $\int_1^\infty \extd \alpha$ part of $Z$ converges (in fact to a bounded oscilliatory function of $L$) but there is a further divergence at $\alpha=0$. The integrand here is a case of
\[ {\extd k\over\extd \alpha}={4\over \alpha^{1\over 2}(8+\alpha)^{3\over 2}},\quad {\extd^m\over\extd\alpha^m}\left({1-e^{ {\imath L\over G}\alpha}\over\alpha}\right)=m!{e^{ {\imath L\over G}\alpha}e_m^{-  {\imath L\over G}\alpha}-1\over (-\alpha)^{m+1}};\quad e_m^x=1+x+{x^2\over 2}+\cdots +{x^m\over m!}.\]
Similarly
\[ \<k_0\>:={\int_{-1}^1\extd k\int_0^L \extd k_0 k_0^2 e^{{\imath \over G}k_0 \alpha(k)}\over \int_{-1}^1\extd k\int_0^L \extd k_0 k_0  e^{{\imath\over G} k_0\alpha(k)}}=-\imath G {\int_0^\infty\extd \alpha{\extd k\over\extd\alpha}{\extd^2\over\extd\alpha^2}\left({1-e^{{\imath L\over G}\alpha }\over \alpha}\right)\over \int_0^\infty\extd \alpha{\extd k\over\extd\alpha}{\extd\over\extd\alpha}\left({1-e^{{\imath L\over G}\alpha }\over \alpha}\right)}=-\imath G \lim_{\alpha\to 0}{ {\extd^2\over\extd\alpha^2}\left({1-e^{{\imath L\over G}\alpha}\over \alpha}\right)\over {\extd\over\extd\alpha}\left({1-e^{{\imath L\over G}\alpha}\over \alpha}\right)} ={2\over 3}L.  \]
Here the $\int_1^\infty \extd \alpha$ part of the numerator converges (in fact to a $L$ times a bounded oscilliatory function of $L$) and there is again a divergence at $\alpha=0$. We then used L'Hopital's rule to find the limit of the ratio of the integrals  as the limit of the ratio of the integrands at the divergent point.  A similar analysis gives in general
\[  \<k_0^m\>:={\int_{-1}^1\extd k\int_0^L \extd k_0 k_0^{m+1}e^{{\imath\over G} k_0\alpha(k)}\over \int_{-1}^1\extd k\int_0^L \extd k_0 k_0  e^{{\imath\over G} k_0\alpha(k)}}=(-\imath G)^m {\int_0^\infty\extd \alpha\, {\extd k\over\extd\alpha}{\extd^{m+1}\over\extd\alpha^{m+1}}\left({1-e^{{\imath L\over G}\alpha}\over \alpha}\right)\over \int_0^\infty\extd \alpha{\extd k\over\extd\alpha}{\extd\over\extd\alpha}\left({1-e^{{\imath L\over G}\alpha}\over \alpha}\right)}={2\over m+2}L^m.\]
We also have $\<k_0^m k^n\>=0 $ for all $n\ge 1$. For odd $n$ this is clear by parity in the original $\int_{-1}^1\extd k$ but it holds for all positive $n$ because if the ratio of integrands has a limit as $\alpha\to 0$, an extra factor $k$ in the numerator makes it tend to zero since  $k=O(\alpha^{1\over 2})$. It also does not change that the numerator integral converges as $\alpha\to\infty$ since $k\sim 1$ for large $\alpha$. In particular, $\<k\>=\<k^2\>=0$.

It follows that 
\[ \<a_{00}\>=\<a_{01}\>=\<k_0(1\pm k)\>={2\over 3}L\]
and one also has $\<a_{00}b_{10}\>=\<a_{00}\>\<b_{10}\>$ etc since the $l_0,l$ integrals operate independently. We use the same cutoff $0\le l_0<L$. It also follows that
\[ \<a_{00}a_{01}\>=\<a_{00}a_{00}\>=\<a_{01}a_{01}\>=\<k_0^2(1\pm k^2)\>={L^2\over 2}\]
which implies for example, a relative uncertainty 
\begin{equation}\label{Deltaa}{\Delta a_{00}\over \<a_{00}\>}={\sqrt{\<a_{00}^2\>-\<a_{00}\>^2}\over \<a_{00}\>}={1\over \sqrt{8}}\end{equation}
for the horizontal edge square-length. Similarly for $b_{10}$ from the other factor for the vertical one.  

The correlation functions themselves have an  infra-red divergence in the same manner as for scalar fields, now appearing as $L\to \infty$ and in principle requiring renormalisation. How to do this in a conventional way is unclear and it may be more appropriate and reasonable (as with the scalar theory) to not renormalise and leave the regulator in place. We can take the operational view that one can  cut-off to $L=3 K_0/2$ to land on any desired $\<a_{00}\>=K_0$, then $\<a_{00}^2\>={9\over 8}K_0^2$ is a calculation for values set at this scale, while the relative $\Delta a/\<a\>$ is independent of this choice of regulator in any case. One might still think of this as some kind of `field renormalisation' to $\hat k_0= {3K_0\over 2L}k_0$ or $\hat a={ 3K_0\over 2L}a$ and similarly for $\hat b$. Then $\<\hat a_{00}\>=\<\hat k_0\>=K_0$ is any desired value resulting from the bare $k_0$  cut off at $L$ while $\<\hat k_0^2(1\pm k^2)\>={9\over 8}K_0^2$  implies the same as (\ref{Deltaa}) for the rescaled  $\hat a_{00}$. However, all we would be doing in practice is  replacing $k_0$ by a new variable $0\le \hat k_0\le {3 K_0\over 2 L}L= 3 K_0/2$ so this just  amounts to the same as setting $L=3 K_0/2$ in the first place. Similarly if one thinks in terms of rescaling the coupling constant $G$.

One can speculate that the constant relative uncertainty (\ref{Deltaa})  is suggestive of some kind of vacuum energy. We also see that a certain amount of geometric structure is necessarily washed out by functional integration in the full quantisation. For example, there is nothing to break the symmetry between $a_{00}$ and $a_{01}$, just as there was no intrinsic scale for $\<a_{00}\>=\<a_{01}\>$ so it had to be convergent or governed by the regulator scale. 

\subsection{Quantisation relative to a Lorentzian rectangular background}

By contrast, it also makes sense to quantise about fixed values and indeed to focus on fluctuations from the rectangular case, which we now do in a relative sense. Thus in the Fourier mode decomposition (\ref{abfou}) of $a,b$ we now fix the average values $k_0,l_0$ as a background rectangle and only quantise relative fluctuations $k,l$ with action (\ref{actionkl}), 
where $\alpha(k)=8 (k^2+k^4+...)$ is approximately Gaussian as for the scalar field on $\Z_2$  in Section~\ref{scalarZ2} (and has its minimum at $k=0$ as expected) but changes as $|k|\to1$ in the gravity case. This is not the usual difference fluctuation from a given background, but fits better with the current computation. In this case, 
\[ Z(k_0,l_0)= 4 k_0 l_0\int_{-1}^1\extd k \int_{-1}^1\extd l \, e^{{\imath\over G} k_0 \alpha(k)- {\imath \over G}l_0 \alpha(l)}\]
where we regard the background rectangle square-lengths $k_0,l_0>0$ as coupling constants and the minus sign in the action comes form the Lorentzian signature. This  converges and we can similarly compute correlations functions from
\[ \<k^2\>={\int_{-1}^1\extd k \int_{-1}^1\extd l e^{{\imath\over G} k_0 \alpha(k)- {\imath\over G} l_0 \alpha(l)} k^2\over 
\int_{-1}^1\extd k \int_{-1}^1\extd l e^{{\imath\over G} k_0 \alpha(k)- {\imath\over G} l_0 \alpha(l)}}\sim {3 G^2\over 128 k_0^2}+{G\over 16 k_0}\imath\]
with the indicated asymptotic form  at large $k_0$ shown in Figure~2. Similarly for $\<l^2\>$ with a conjugate answer. From these we have
\[ \<a_{00}\>=\<a_{01}\>=\<k_0(1\pm k)\>=k_0\]
\[  \<a_{00}^2\>=\<a_{01}^2\>=k_0^2(1+ \<k^2\>),\quad \<a_{00}a_{01}\>=k_0^2(1-\<k^2\>)\]
and similarly for $b$. For the same reasons as before, we also have $\<a_{00}b_{10}\>=k_0l_0$ etc. (and similarly for  $a,b$ at any other points). In short, the edge square-lengths $a$ and $b$ behave independently and each is similar to a scalar field on a 2-point graph, but $\<k^2\>$ only asymptotes as $k_0\to \infty$ to the constant imaginary value $\imath/(16 k_0)$ from (\ref{vevk2scalar}) for the scalar case on $\Z_2$. This justifies the view of large $k_0$ as some kind of large scale or low energy limit. At the other limit we have, by contrast,
\[ \lim_{k_0\to 0} \<k^2\>= {1\over 3}.\]
The physical meaning of this is unclear due to our model having only four points but  we have one interpretation in this limit  as a relative edge-length uncertainty $\Delta a_{00}/\<a_{00}\>=\sqrt{\<k^2\>}=1/\sqrt{3}$ similar to our previous (\ref{Deltaa}). 

\begin{figure}
\[ \includegraphics[scale=.8]{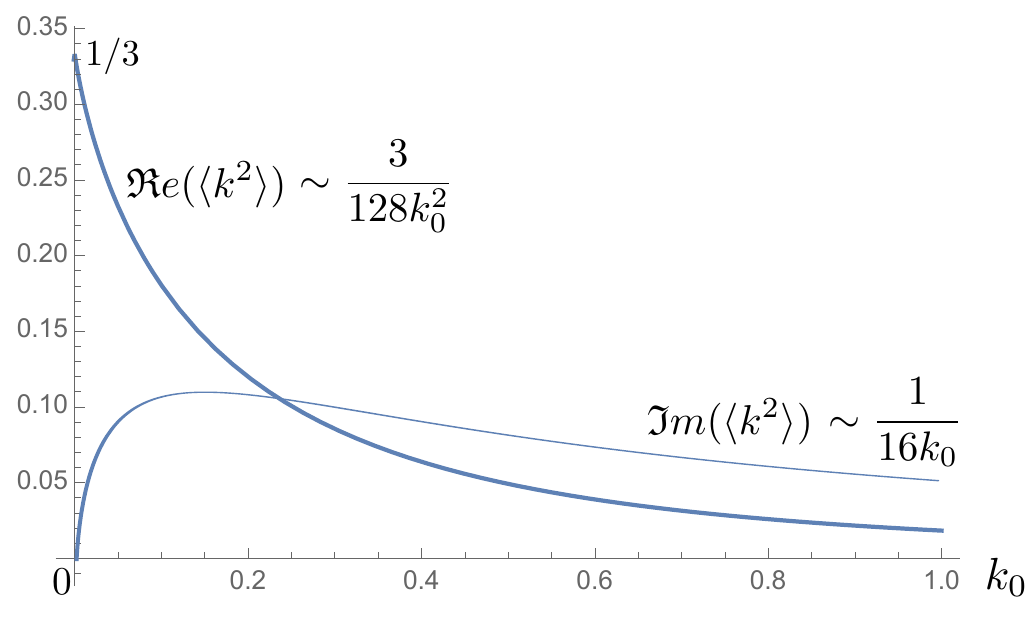}\]
\caption{Expectation value $\<k^2\>$ for relative quantum metric fluctuations on a background Lorentzian rectangle with sides $k_0,l_0$ at $G=1$. Compare with  $\<k^2\>={\imath\over 16 k_0}$ for the scalar field case from (\ref{vevk2scalar}). \label{fig2}}\end{figure}

\section{Concluding remarks}

We have seen that a universe of four points with a quadrilateral differential structure has a natural quantum Riemannian geometry which leads to a plausible, if not completely canonical, Einstein-Hilbert action, which in turn can be quantised in a functional  integral approach. Choices for the action were a lifting map $i$ needed to make a trace to define the Ricci tensor (we took the obvious antisymmetric lift) and the `measure' $\mu$ for integrating the action over the four points (we took $|\det(g)|=ab$ as this rendered the action independent of the freedom $q$ in the Levi-Civita connection). We also chose the horizontal edges to actually be assigned negative values, which we called the Lorentzian case. 

We  first quantised a free scalar field on two and four points with the expected Gaussian form that now depends on the modulus one parameter $q$ in the QLC, for which we focused on the real case of $q=\pm1$. We saw in Section~\ref{scalarcurved} that this freedom is needed to implement the symmetry of the square which in the Euclidean version flips horizontal and vertical.  For the quantum gravity theory, we found that the edge square-lengths $a_{00},a_{01}$ on horizontal edges and  $b_{10}, b_{11}$ on vertical edges proceeded independently. Thus,  we can think of each horizontal edge as a `point' and $a$ as a real valued scalar field with values on the bottom or top edge (see Figure~\ref{fig1}) together with a non-quadratic action $k_0\alpha(k)$ in terms of the relative fluctuation $k$ and the average horizontal edge square-length $k_0$. Similarly for the vertical theory based on the variables $l,l_0$ and the action entering with an opposite sign (so with conjugate results). 

We found that both the full theory where $k,k_0$ and $l,l_0$ are quantised and the massless scalar theory have expected IR divergence easily regulated by a square-length scale $L$ and mass $m$ respectively.  We also  computed correlation functions. For scalar fields they have an expected imaginary form but for the fully fluctuating quantum gravity case the correlation functions were real with an interpretation as a constant nonzero relative uncertainty in the quantum edge square-lengths. We argued that there may be no need to remove the regulator (just as there is no need to work with massless scalar fields). We also looked at an intermediate version of the quantum gravity theory where only the relative fluctuations $k,l$ are quantised and $k_0,l_0$ are treated as a background rectangular geometry.  We found that this interpolates between the full quantisation as $k_0,l_0\to 0$ and two free scalar theories on $\Z_2$ as $k_0,l_0\to\infty$. 

There are several questions which a theory of quantum gravity should be able to answer and to which even our baby model of four points could potentially give insight  in future work. One direction is the hint from the fixed relative uncertainty result of a background `dark energy'. Another could be to explore the expected link between gravity and entropy in its various forms and possibly to compute Hawking radiation via the generic curved background Laplacian of Section~\ref{scalarcurved}. At a technical level one may  also be able to look at Einstein's equation in the combined quantisation of both scalar fields and gravity.  In such a theory, since the QLC is not uniquely determined by the metric and the scalar Laplacian is sensitive to the ambiguity (which could be viewed as changing signature), we should really sum over this freedom in the QLC's. These matters will be considered in a sequel. 

Moreover, the methods of the paper can be applied in principle to any graph. The quantum Riemannian geometry for the triangle case is solved in \cite{BegMa:book} and is not too interesting for quantum gravity, but there are many other graphs and, possibly, infinite lattices that one could apply the same methods to. The latter would be necessary if one wanted to have some kind of `continuum limit' rather than our point of view that a discrete set has a geometry all by itself. The physical interpretation is generally clearer when there is a continuum limit, whereas for a finite system more experience will be needed possibly including ideas from quantum information.  The infinite lattice case is, however,  hard to solve for  general metrics due to the non-linear nature of the QLC conditions. Similarly for a finite graph that is not a Cayley graph on a finite group, one does not then have Fourier transform making it again hard to solve for the QLCs. Both cases can and should  ultimately be compared with lattice quantum gravity results such as in \cite{Lat} although at present the methodologies are very different. We also note that the formalism of quantum Riemannian geometry works over any field, opening another front over small fields\cite{MaPac} where QLCs may be more directly computable. 

The methods of the paper can of course be applied to other algebras including noncommutative ones, but the physical interpretation is likely to be even less clear. There is, for example, a rich moduli of differential structures and metrics on $M_2(\C)$, see \cite{BegMa3}\cite{BegMa:book} but QLCs have so far only been determined for some specific metrics.

\end{document}